%% file: main_R2.tex
\documentclass[journal]{IEEEtran}

\usepackage{amsmath,amssymb}
\usepackage{graphicx}
\usepackage{booktabs}
\usepackage{multirow}
\usepackage{array}
\usepackage{siunitx}
\usepackage{url}
\usepackage[hidelinks]{hyperref}

\newcommand{\SEVF}{SE(VFib)}
\newcommand{\SEVT}{SE(VTach)}
\newcommand{\SPN}{SP(NSR)}
\newcommand{\SPA}{SP(Asys)}
\newcommand{\SPO}{SP(Oth)}
\newcommand{\SPOA}{SP(Oth+Asys)}

\newcommand{\figCascadeW}{0.70\columnwidth}   
\newcommand{\figArchW}{1.00\columnwidth}      
\newcommand{\figRhythmW}{1.00\columnwidth}    
\newcommand{\figAmplitudeW}{1.00\columnwidth} 

\begin{document}

\title{Sensor-Free Shockable Rhythm Classification During CPR Using the Existing ECG Channel of Public-Access Automated External Defibrillators}


\author{Jaechan~Lim,
        David~Hicks,
        Matt~Valentine,
        and~Ki~H.~Chon,~\IEEEmembership{Fellow,~IEEE}
\thanks{This work was supported by the National Heart, Lung, and Blood
Institute of the National Institutes of Health under Award No. 1R01HL173935.
\emph{(Corresponding author: Jaechan Lim.)}}
\thanks{J. Lim and K. H. Chon are with the Department of Biomedical
Engineering, University of Connecticut, Storrs, CT 06269 USA
(e-mail: jaechan.lim@uconn.edu; ki.chon@uconn.edu).}
\thanks{D. Hicks and M. Valentine are with Defibtech LLC, Guilford,
CT 06437 USA. (e-mail: dhicks@defibtech.com; mvalentine@defibtech.com)}}


\maketitle

\begin{abstract} 
Automated external defibrillators (AEDs) are public-access medical devices deployed in workplaces, schools, and homes and intended for use by lay rescuers, yet they must pause chest compressions to analyze the rhythm because compressions corrupt the electrocardiogram (ECG). Removing this pause requires a method that works with the hardware already present in the device, fits within the computational budget of its processor, and meets the rhythm-recognition performance required for defibrillator approval. Existing approaches rely either on an auxiliary sensor, which entails a hardware redesign, or on an additional network that reconstructs the ECG before classification. This work classifies the compression-contaminated ECG directly, using only the ECG channel that an AED already acquires. A cascade of three compact convolutional networks trained with different class ratios verifies each decision with a second network biased toward the proposed outcome, and disagreement between the two yields an indeterminate outcome. Evaluated on an independent manufacturer-provided cohort of 464 subjects over 15 train-validation partitions, the cascade satisfies both adjudicable American Heart Association criteria under all six noise conditions at an eight-second window and a compression signal-to-artifact ratio of 0~dB, with sinus-rhythm specificity of 99.20--99.43\% and specificity of 95.48--96.20\% for the other non-shockable rhythms, while leaving only 1.07--1.48\% of segments indeterminate. The criteria are not met at $-3$~dB. Single networks matched in parameter count and inference cost do not reproduce this performance. Each decision evaluates two of the three networks and requires 6.68~ms on a single thread of a general-purpose processor, and eight-bit quantization reduces the model to 1.45~MiB without changing any compliance verdict. These results show that rhythm analysis during ongoing CPR can be implemented using the ECG hardware already present in public-access AEDs, avoiding the need for auxiliary sensing hardware while remaining subject to validation on the target device. 
\end{abstract}


\input{01_introduction_R2}
\input{02_related_work_R2}
\input{03_data_R2}

\input{04_method_R2}
\input{05_experimental_setup_R2}
\input{06_results_R2}

\input{06a_deployment_R2}

\input{07_discussion_R2}

\input{08_conclusion_R2}

\bibliographystyle{IEEEtran}
\bibliography{ref_cpr_cnn}

\end{document}

%% file: 01_introduction_R2.tex
\section{Introduction}
\label{sec:intro}

\begin{figure*}[!t]
\centering
\includegraphics[width=\textwidth]{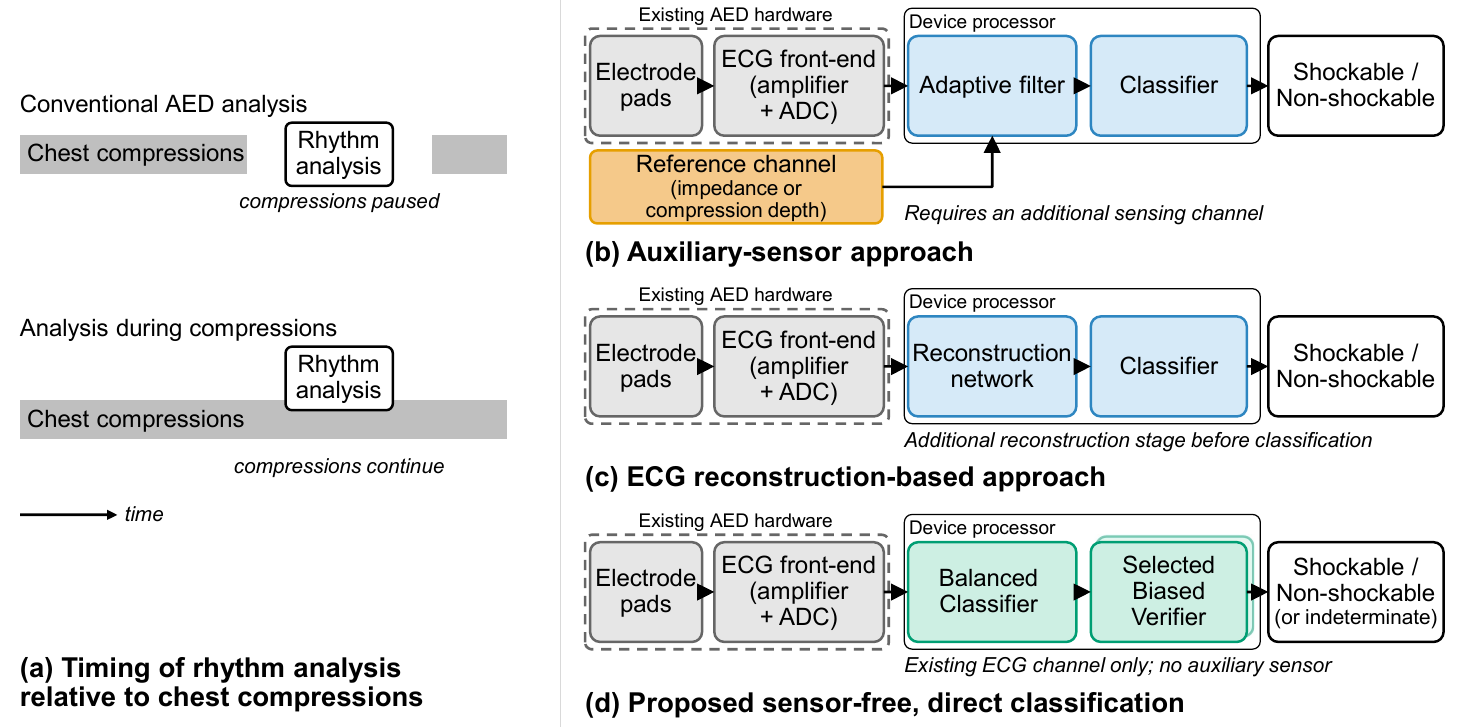}
\caption{Rhythm analysis during chest compressions in automated external defibrillators. (a) Timing of rhythm analysis relative to chest compressions. Conventional rhythm analysis is typically performed during a pause in compressions, whereas the approaches considered in this work aim to analyze the rhythm while compressions continue. Only the timing of analysis is illustrated; device prompts, shock delivery, and interruption schedules are not shown. (b) Methods that estimate the compression artifact from an auxiliary reference require a sensing channel beyond the ECG. (c) ECG-only reconstruction methods add a reconstruction stage before classification. (d) The proposed cascade classifies the compression-contaminated ECG directly through the existing ECG acquisition pathway without requiring an auxiliary sensing channel. The dashed boundary in (b)--(d) marks hardware already present in the AED.}
\label{fig:system}
\end{figure*}

Reliable rhythm classification during cardiopulmonary resuscitation (CPR) is challenging because the electrocardiogram (ECG) needed for decision making is contaminated by the therapy intended to preserve life. Survival after out-of-hospital cardiac arrest depends on both the rapid identification of shockable rhythms and the continuous delivery of chest compressions \cite{cummins1991chain, waalewijn2001survival}, yet compressions produce an electrical artifact large enough to obscure the underlying ECG. Automated external defibrillators (AEDs) therefore interrupt compressions before analyzing the rhythm (Fig.~\ref{fig:system}(a)), even though longer pauses before defibrillation are associated with lower survival to hospital discharge \cite{cheskes2011perishock, waalewijn2002prevention}. This trade-off must be resolved not in a hospital but in public-access medical devices deployed in workplaces, schools, and homes and intended for use by lay rescuers who follow the instructions issued by the device \cite{hazinski2005lay, fda_aed_2024}. 

Because such devices may remain installed for years between uses and must analyze the rhythm without a trained professional to interpret the trace, a method for analysis during compressions is useful only if it fits the product in which it will operate. It must use the hardware already present in the device, since each additional sensor increases hardware cost and turns an algorithmic improvement into a hardware redesign. It must reach a decision on the processor inside the AED within the timescale of the compression cycle rather than relying on workstation-class computing resources, a constraint that has likewise shaped cardiac signal analysis on portable and wearable consumer devices \cite{chen2023srecg}. It must also meet the rhythm-recognition performance required for defibrillator approval, which is specified by the American Heart Association (AHA) and reproduced in IEC 60601-2-4 \cite{kerber1997automatic, iec60601_2_4}. A method satisfying these requirements would make rhythm analysis during ongoing CPR a function of signal-processing software rather than sensing hardware while preserving the existing ECG acquisition pathway, subject to regulatory review and validation on the target platform.
Fig.~\ref{fig:system}(b)--(d) contrasts this pathway with the two established strategies reviewed below.

Eliminating the pause requires rhythm classification to be performed while chest compressions continue. The challenge is not only the magnitude of the compression artifact but also its spectral overlap with the underlying rhythm. Compressions delivered at 100 to 120 per minute generate energy near 1.7 to 2~Hz and at their harmonics, substantially overlapping the dominant frequency content of ventricular fibrillation. Consequently, filtering the artifact inevitably removes some of the signal of interest as well.

Two broad strategies have been developed to address CPR artifacts. The first uses an auxiliary measurement, such as transthoracic impedance or compression depth, to estimate and subtract the artifact \cite{irusta2009least,ruiz2010cardiopulmonary}. This approach has been adopted commercially, but it depends on a sensing channel that is unavailable in many existing AEDs.

The second strategy relies on the ECG alone. Instead of estimating the artifact directly, it reconstructs a cleaner ECG waveform and then classifies the reconstructed signal \cite{rheinberger2008removal,gong2014removal,li2023transfer,nejad2024enhancing}. Because no additional hardware is required, this approach has attracted substantial research interest. Its performance, however, depends on reconstruction fidelity, since any distortion introduced during reconstruction is inherited by the downstream classifier. Reconstruction also introduces a second network into the processing pipeline, adding to the computational load on the device processor even when reconstruction does not affect the final decision.

Our earlier work highlighted this dependency directly \cite{lim2026cascade}. When a commercial classifier in a reconstruction-based pipeline was replaced with a convolutional neural network trained on a larger and more diverse dataset, distortions introduced during reconstruction became apparent. Asystole was reconstructed with organized-looking waveforms, while atrial flutter exhibited oscillations larger than those present in the clean ECG. As a result, specificity deteriorated for both rhythms. Although the reconstruction remained adequate for the classifier with which it was originally developed, its performance did not transfer unchanged to the new classifier. These findings indicate that artifact removal and classification cannot always be treated as independent stages of the pipeline, because reconstruction errors may affect different classifiers in different ways.

A third approach dispenses with reconstruction altogether. Instead of first estimating a clean electrocardiogram, a classifier assigns a shockable or non-shockable label directly from the compression-contaminated signal \cite{hajeb2021deep,krasteva2023deep,jekova2021optimization}. The challenge is that rhythm-classification requirements are inherently asymmetric. The AHA specifies separate performance targets for five rhythm classes \cite{kerber1997automatic}. For example, specificity for sinus rhythm must exceed 99 percent, whereas sensitivity for coarse ventricular fibrillation must exceed 90 percent. A single classifier trained with a single objective is not explicitly optimized for this asymmetry. While increasing the decision threshold can improve specificity, the resulting loss in sensitivity may make it difficult to satisfy both requirements simultaneously. This motivates an architecture that explicitly addresses the asymmetry between shockable and non-shockable rhythm recognition.

This work addresses that asymmetry through a cascade of three convolutional networks that differ only in the class ratios used during training. A balanced network first generates a shockable or non-shockable classification, which is then verified by a second network biased toward the proposed outcome. Shockable classifications are verified by a network trained to favor shockable rhythms, whereas non-shockable classifications are verified by a network trained to favor non-shockable rhythms. When the verification network fails to confirm the initial decision, the evidence supporting that decision is considered insufficient and the segment is assigned to an indeterminate category. Although the cascade contains three networks, each decision evaluates only two of them, resulting in computational cost comparable to that of a single network of similar size.

The credibility of a method of this kind depends not only on its performance but also on the rigor of its evaluation. Because prior studies report training-test separation inconsistently, the evaluation protocol is stated explicitly. Training and testing are separated simultaneously by patient, compression artifact, and noise segment, and the patient partition is regenerated 15 times. As a result, the reported variation reflects sensitivity to the training population rather than dependence on a single train-test split. In addition, indeterminate outcomes are counted rather than excluded. The reported performance therefore incorporates the cost of uncertainty instead of removing difficult cases from the evaluation.

This work makes four main contributions. First, it demonstrates shockable-rhythm classification during ongoing CPR that satisfies the adjudicable AHA criteria without an artifact-removal stage, eliminating the need for a reconstruction network in the decision pipeline. Second, it achieves this using only the ECG channel already acquired by an AED, without a transthoracic impedance channel, compression-depth sensor, or other auxiliary measurement. Third, it evaluates the proposed method on an independently collected manufacturer-provided AED cohort of 464 subjects, with training and testing separated by patient, compression artifact, and noise segment across 15 independently generated partitions. Finally, it characterizes the computational cost of the method through arithmetic, memory footprint, and single-thread latency measured on a general-purpose processor, together with the effect of eight-bit quantization on AHA compliance. These results provide an initial assessment of its suitability for deployment on AED processors.

The remainder of the paper is organized as follows. Section~\ref{sec:related} reviews prior work, and Sections~\ref{sec:data} and \ref{sec:method} describe the data and the cascade. Section~\ref{sec:setup} states the evaluation protocol, Section~\ref{sec:results} reports performance, and Section~\ref{sec:deployment} reports the computational cost relevant to deployment. Section~\ref{sec:discussion} places the result among published methods, discusses its implications for public-access AED platforms, and states the limitations. 

%% file: 02_related_work_R2.tex
\section{Related Work}
\label{sec:related}

Among the three categories introduced in Section~\ref{sec:intro}, the present work belongs to the group of methods that classify compression-contaminated electrocardiograms directly, without an explicit artifact-removal stage. Deep neural networks have been trained on human recordings mixed with compression artifacts \cite{hajeb2021deep}, and end-to-end optimization has been applied to rhythms recorded during out-of-hospital resuscitation \cite{jekova2021optimization}. Feature-based approaches have also been reported for piston-driven compressions \cite{isasi2019machine}, and the commercial AWC algorithm likewise operates on the electrocardiogram alone \cite{didon2024clinical}.

These approaches avoid the dependence on reconstruction fidelity that characterizes artifact-removal pipelines. However, they do not all make the same assumptions about the available ECG signal. For example, the sliding-window method of \cite{krasteva2023deep} requires two to three seconds of relatively clean electrocardiogram within the analysis window. Although this requirement is less restrictive than the use of an auxiliary sensing channel, it still imposes constraints on the signal presented to the classifier.

Commercial systems differ substantially in how they handle uncertain classifications. The cprINSIGHT algorithm reports no decision for approximately 30 percent of clinical analyses and computes its published performance only on the remaining analyses \cite{degraaf2021analyzing}. In contrast, the AWC algorithm avoids an explicit indeterminate category. When a shockable rhythm is detected during compressions, the device pauses CPR and performs a confirmation analysis on a cleaner ECG signal. This additional verification step occurs in 27.7 percent of analyses \cite{didon2024clinical}.

These approaches address uncertainty in fundamentally different ways. One excludes a subset of analyses from the reported performance metrics, whereas the other seeks additional evidence before issuing a final decision. Both are valid design choices, but neither produces performance estimates that are directly comparable with rates computed over all analyzed segments.

Selective classification treats abstention as a decision in its own right and formalizes the trade-off between coverage and risk \cite{geifman2017selective}. Because a classifier can always improve its apparent performance by declining to classify difficult cases, reported error rates depend not only on classification accuracy but also on the fraction of segments that receive a decision. Consequently, performance measured only on answered segments cannot be compared fairly across methods operating at different levels of coverage.

The proposed cascade is also a selective classifier, but its reported performance is not limited to segments receiving a definitive classification. Abstention arises from disagreement between two networks rather than from a confidence threshold applied to a single network. Under the intention-to-diagnose convention, indeterminate outcomes remain part of the evaluation and are not removed from the denominator. Consequently, the reported performance incorporates the cost of uncertainty rather than excluding difficult cases from analysis. The resulting rates are therefore directly comparable with those of classifiers that never abstain. Section~\ref{sec:results} further compares the proposed abstention mechanism with alternative abstention strategies under matched coverage.

%% file: 03_data_R2.tex
\section{Data}
\label{sec:data}

This study draws on three independent data sources, each partitioned to maintain strict separation between training and testing. PhysioNet recordings provide the training and validation data, AED validation records provide the test cohort, and compression artifacts and clinical noise are drawn from separately partitioned pools.

\subsection{Training and Validation Pool}

Training data are drawn from six PhysioNet databases \cite{goldberger2000physiobank}. The MIT-BIH Arrhythmia Database \cite{moody2001impact} provides sinus rhythm and most non-shockable rhythms, whereas the MIT-BIH Atrial Fibrillation Database \cite{moody1983afdb} provides atrial fibrillation and atrial flutter. Ventricular fibrillation and ventricular tachycardia are obtained primarily from the MIT-BIH Malignant Ventricular Arrhythmia Database \cite{greenwald1986development} and the Creighton University Ventricular Tachyarrhythmia Database \cite{nolle1986crei}. The MIT-BIH Supraventricular Arrhythmia Database \cite{greenwald1990improved} and the St.\ Petersburg INCART 12-lead Arrhythmia Database \cite{incartdb} provide recordings from which fast narrow-complex rhythm segments are extracted. A seventh database, the MIT-BIH Sudden Cardiac Death Holter Database, was evaluated but excluded because it contains only detector-generated output in its \texttt{.ari} files and does not provide expert rhythm annotations.

All signals are resampled to 125~Hz using a polyphase resampling method with a finite-impulse-response anti-aliasing filter applied before decimation. The resampled signals are then high-pass filtered at 0.5~Hz using a third-order Butterworth filter applied in a forward-backward manner and are subsequently mean-subtracted. This preprocessing ensures that aliasing is controlled during resampling rather than assessed retrospectively after conversion.

Two curation decisions were required before rhythm labels could be accepted for training. The first concerns the annotation string \texttt{(N}. In the MIT-BIH Arrhythmia Database, this label denotes sinus rhythm, whereas in several other databases it functions as a catch-all category that may also include agonal rhythm and signal dropout. To avoid label inconsistency, sinus-rhythm segments are therefore drawn only from the MIT-BIH Arrhythmia Database and are excluded from recordings containing atrial fibrillation or pacing.

The second decision concerns asystole. Candidate asystole segments are screened using an amplitude criterion because many labeled examples are more consistent with severe bradycardia or pulseless electrical activity than with true electrical standstill. At the eight-second window length, this screening procedure retains 45 of the 92 candidate segments.

The resulting dataset contains 8{,}559 segments from 146 recordings spanning 21 rhythm classes, including 1{,}484 shockable segments. Thirteen of the 21 rhythm classes are absent from the test cohort but are retained to expose the model to a broader range of ECG morphologies during training and thereby improve generalization.
\subsection{Test Cohort}
\label{sec:testcohort}

The test cohort consists of 464 subjects, each contributing a single 14-second ECG recording sampled at 125~Hz from one lead. The recordings originate from AED validation data provided by Defibtech LLC, and all rhythm labels were supplied with the recordings rather than assigned by this study. The original dataset contains 467 subjects; three subjects with heart block were excluded so that the cohort matches the population used in the companion study and previous analyses of the same records \cite{lim2026cascade, nejad2024enhancing}. The resulting cohort composition is summarized in Table~\ref{tab:testcohort}.

This population does not appear in any PhysioNet database. Consequently, patient-level leakage between the training and test sets is eliminated by design. Unlike a patient-level split performed within a single database, this arrangement preserves access to all available rare rhythms for training while still providing an independent test cohort. It also enables repeated record-level partitioning of the training pool without sacrificing rare rhythm classes to a fixed test partition.

\begin{table}[!t]
\caption{Composition of the test cohort. The ``Training counterpart'' column indicates whether the training pool contains a rhythm category with the same label. The ``Other non-shockable'' category includes subjects whose source records specify only the shock decision rather than a more detailed rhythm annotation \cite{nejad2024enhancing}. According to the amplitude and rate definitions in the AHA recommendations \cite{kerber1997automatic}, all 41 ventricular-fibrillation subjects satisfy the definition of coarse ventricular fibrillation, and 29 of the 31 ventricular-tachycardia subjects satisfy the definition of rapid ventricular tachycardia.}
\label{tab:testcohort}
\centering
\setlength{\tabcolsep}{2pt}
\scriptsize
\begin{tabular}{@{}l>{\raggedright\arraybackslash}p{0.40\columnwidth}rc@{}}
\toprule
Class & Rhythm & Subjects & Training counterpart \\
\midrule
Shockable & Ventricular fibrillation & 41 & yes \\
          & Ventricular tachycardia  & 31 & yes \\
\midrule
Non-shockable & Sinus rhythm             & 169 & yes \\
              & Supraventricular tachycardia   &  47 & yes \\
              & Atrial fibrillation      &  33 & yes \\
              & Atrial flutter           &  13 & yes \\
              & Other non-shockable      &  55 & no \\
              & Premature ventricular complexes &  37 & no \\
              & Rapid supraventricular tachycardia &  21 & no \\
              & Asystole                 &   9 & yes \\
              & Sinus bradycardia        &   8 & yes \\
\midrule
Total &  & 464 & \\
\bottomrule
\end{tabular}
\end{table}

Of the 392 non-shockable subjects, 113 (28.8\%) belong to rhythm categories that are not explicitly included in the training pool. To distinguish performance on included and non-included rhythms, specificity is reported separately for three groups: rhythms with direct training counterparts, supraventricular tachycardia whose training counterpart differs substantially in spectral characteristics, and rhythms with no corresponding training counterpart.

The test recordings are filtered and mean-subtracted using the same preprocessing pipeline applied to the training data. No resampling is performed, and shorter analysis windows are obtained through center cropping. No quality screening is applied, since excluding test recordings according to criteria derived from the training data would effectively influence which cases are evaluated.

At the eight-second window length, the median signal standard deviation across the cohort is 0.2832~mV. The lower end of the distribution is dominated by asystole, whose median standard deviation is 0.0055~mV compared with 0.2446~mV for sinus rhythm. This approximately 45-fold difference in signal amplitude motivates the amplitude input introduced in Section~\ref{sec:amplitude}.

\subsection{Compression Artifacts and Noise}

The CPR-artifact pool consists of 52 recordings, each 14 seconds long and sampled at 125~Hz, acquired during chest compressions delivered to patients in cardiac arrest by different rescuers using Defibtech automated external defibrillators \cite{hajeb2022enhancing, nejad2024enhancing}. Because the recordings were collected during asystole, each trace contains compression artifact with minimal underlying cardiac activity. This characteristic is exploited by the augmentation strategy described in Section~\ref{sec:augmentation}. No two recordings originate from the same patient.

Twelve recordings are excluded because they contain compression pauses longer than one second. The remaining 40 recordings are partitioned into 32 training, 4 validation, and 4 test recordings, with the artifact pool repartitioned for each train-validation split.

Electrode motion, baseline wander, and muscle artifact are obtained from the MIT-BIH Noise Stress Test Database \cite{moody1984noise}, which provides 256 fourteen-second segments for each noise type. The noise pool is partitioned temporally rather than by recording channel. For each channel, the first half of the recording forms the training pool and the second half forms the test pool, ensuring that both leads remain included in both partitions.

A channel-based partition was not used because the two leads differ substantially in gain, by a factor of 6.5 for electrode motion. Under such a split, any loss of performance would be difficult to attribute uniquely to an unseen noise segment because it would also be confounded by differences between recording channels. The temporal partition avoids this confounding while still maintaining complete separation between the training and test noise segments.

Validation uses the training noise pool rather than a dedicated validation-noise partition. The purpose of validation is to monitor training and select the operating threshold rather than to estimate generalization performance. Reserving additional noise solely for validation would effectively create a second test set and could bias threshold selection toward optimistic estimates of performance.

\subsection{Partitioning}

All segments originating from the same recording are assigned to either the training or validation partition, with no recording appearing in both. The recordings are divided into five groups, and the grouping procedure is repeated three times, yielding a total of 15 train-validation splits. Assignment is performed using a greedy strategy that balances both the number of shockable segments and the total number of segments, while allowing a 15\% tolerance so that the three repetitions produce distinct partitions. The mean best-matching Jaccard index between repetitions ranges from 0.156 to 0.270, indicating that the resulting partitions are meaningfully different from one another.

Training-test independence is enforced along three dimensions. First, patient independence is guaranteed because the training and test cohorts originate from different institutions. Second, the compression-artifact pool is partitioned into 32 training, 4 validation, and 4 test recordings with no overlap between partitions. Third, the noise pool is divided temporally so that no noise segment appears in more than one partition.

\subsection{Ethical Considerations}
This study utilized de-identified human subject data from publicly available PhysioNet databases and independent AED validation recordings provided by Defibtech LLC. All data were anonymized prior to analysis. Because the study involved only secondary analysis of existing de-identified data, additional institutional review board review was not required.

%% file: 04_method_R2.tex
\section{Method}
\label{sec:method}

\subsection{Structure and Decision Rule} 
The proposed cascade consists of three identical convolutional networks trained on the same data but with different class ratios. The balanced network is trained using equal numbers of shockable and non-shockable samples in each epoch. The sensitivity-biased network uses an 80:20 shockable-to-non-shockable ratio, whereas the specificity-biased network uses the reverse ratio of 20:80. Apart from these class-ratio differences, all aspects of the training procedure and network architecture are identical. 

\begin{figure}[!t]
\centering
\includegraphics[width=\figCascadeW]{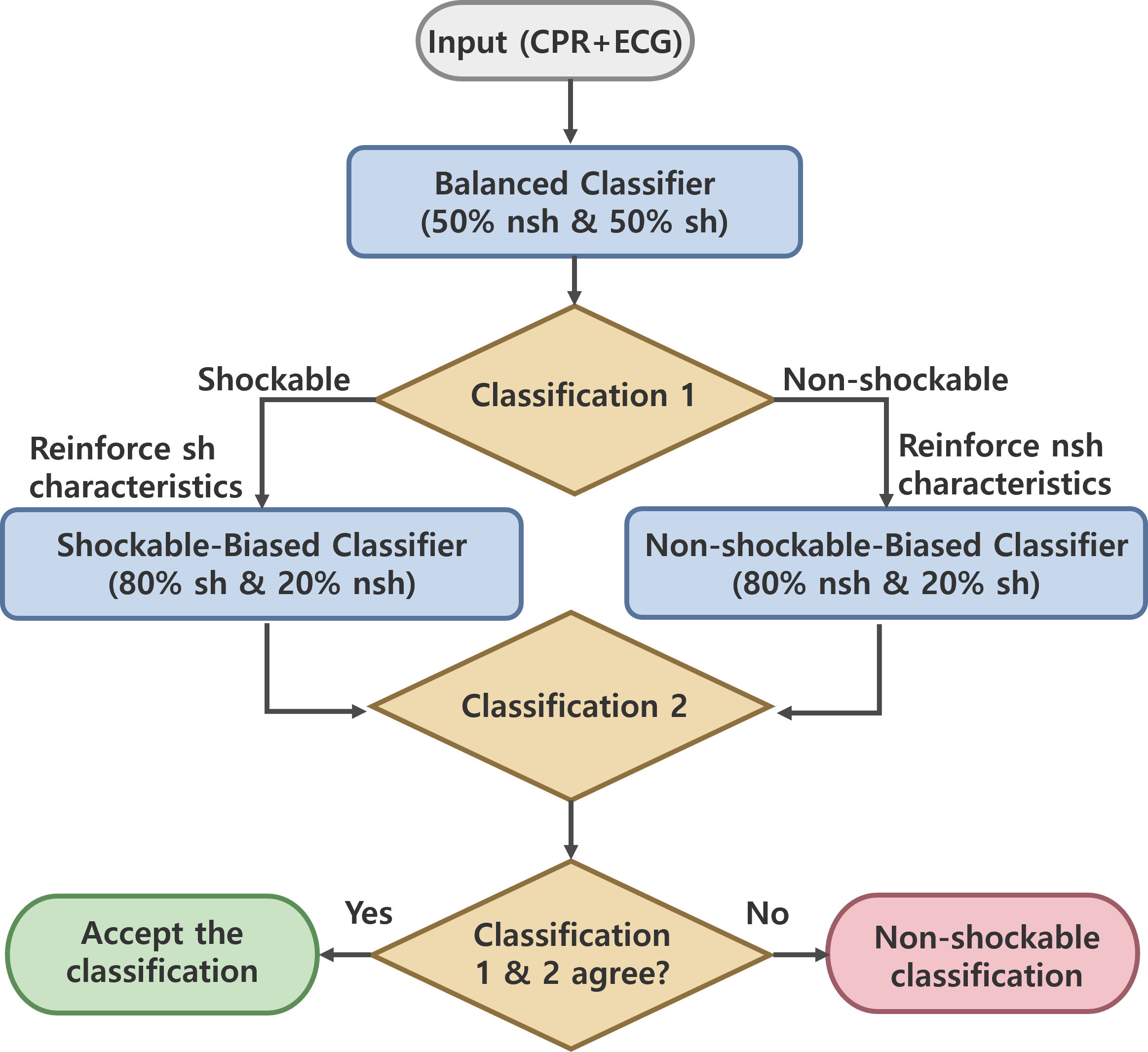}\\[1pt]
(a)\\[4pt]
\includegraphics[width=\figArchW]{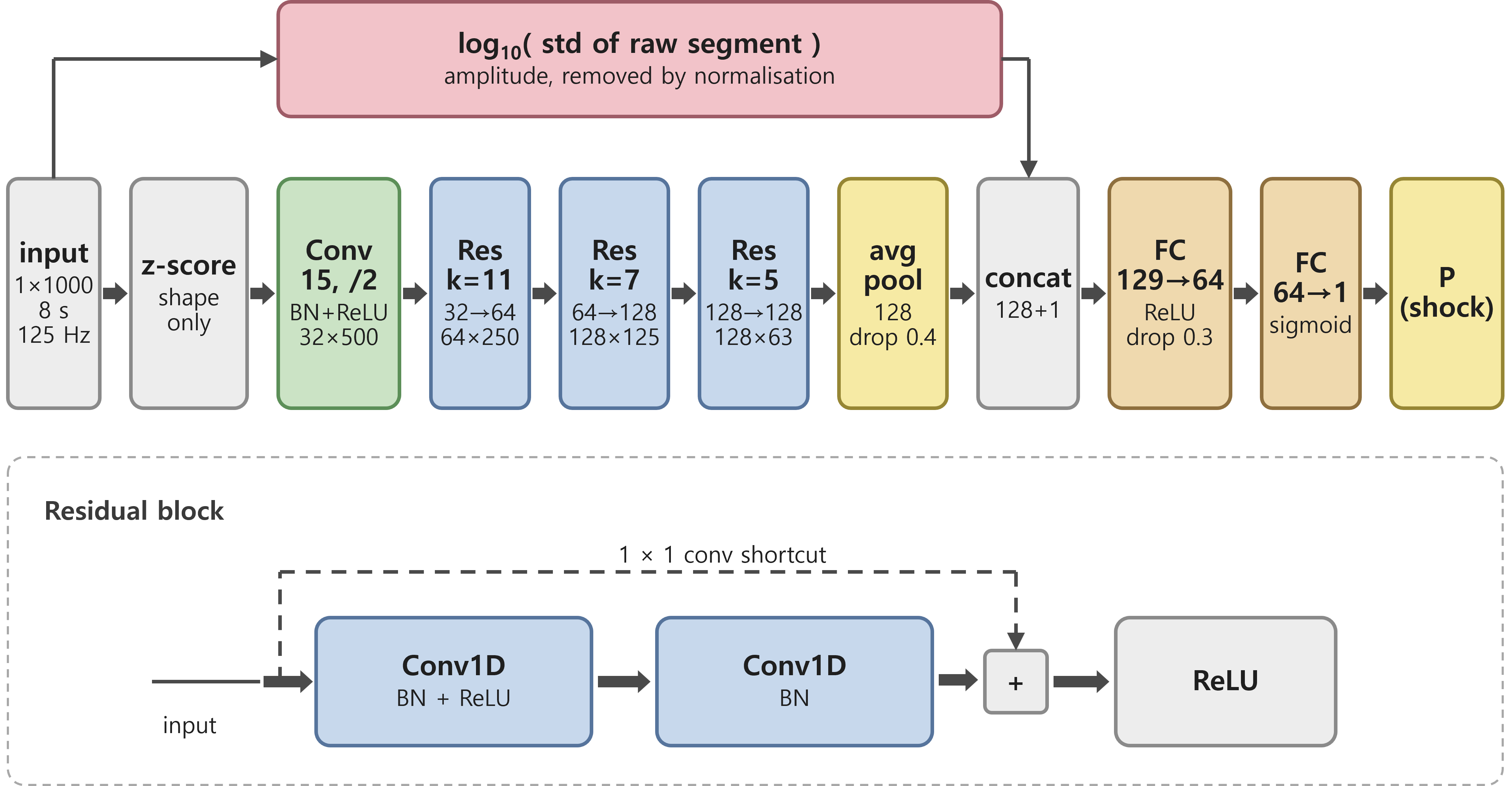}\\[1pt]
(b)
\caption{(a) Decision process of the proposed cascade. The balanced network generates an initial shockable or non-shockable classification, which is subsequently verified by the corresponding biased network. Disagreement between the two stages results in an indeterminate outcome. (b) Architecture of the proposed network. The amplitude branch receives the signal before normalization, preserving amplitude information that would otherwise be removed. Its scalar output is concatenated with the pooled features. Tensor shapes correspond to the eight-second analysis window.}
\label{fig:arch}
\end{figure}

Let $p_{\mathrm{bal}}$, $p_{\mathrm{sh}}$, and $p_{\mathrm{nsh}}$ denote the predicted probabilities produced by the balanced, sensitivity-biased, and specificity-biased networks, respectively, and let $t$ denote the operating threshold. The balanced network first generates a binary classification,
\begin{equation} 
a = \mathbf{1}\!\left[p_{\mathrm{bal}} \geq t\right], 
\end{equation} 
and a confirmation decision $c$, generated by the biased network corresponding to the proposed classification, 
\begin{equation} 
c = \begin{cases} \mathbf{1}\!\left[p_{\mathrm{sh}} \geq t\right], & a = 1,\\[2pt] \mathbf{1}\!\left[p_{\mathrm{nsh}} \geq t\right], & a = 0. \end{cases} 
\end{equation} 
The segment is declared indeterminate when the two decisions disagree, 
\begin{equation} 
\mathrm{abstain} = \mathbf{1}\!\left[a \neq c\right], 
\end{equation} 
and the initial classification is retained whenever the two networks agree. A shockable classification is verified by the network trained to favor shockable outcomes, whereas a non-shockable classification is verified by the network trained to favor non-shockable outcomes. Failure of the verification stage indicates that the evidence supporting the proposed classification is weak, even for the classifier most biased toward that outcome. Fig.~\ref{fig:arch}(a) illustrates the resulting decision process.

An indeterminate outcome is not treated as a third class. Its contribution to the reported performance metrics is defined in Section~\ref{sec:metrics}.

\subsection{Network}

Each network receives a single-channel input of 1{,}000 samples, corresponding to an eight-second ECG segment sampled at 125~Hz, and produces a single output logit. The architecture consists of a stem convolution followed by three residual blocks with kernel sizes of 11, 7, and 5. Each block uses a stride of 2, increasing the number of channels from 32 to 128 while progressively reducing the sequence length. Global average pooling, dropout, and two fully connected layers generate the final prediction. The overall architecture is illustrated in Fig.~\ref{fig:arch}(b).

Strided convolutions are used instead of dilation to keep the network compact and computationally efficient. Because each residual block halves the sequence length, the deepest block operates on 125 samples and produces an output length of 63 rather than the original 1{,}000 samples. Each network contains 441{,}921 parameters, giving a total of 1{,}325{,}763 parameters for the full cascade.

\subsection{The Amplitude Input}
\label{sec:amplitude}

Each segment is normalized to zero mean and unit variance before being presented to the network. Because signal gain varies substantially across the training databases, normalization allows a single model to operate across all recordings without database-specific scaling. However, normalization also removes amplitude information that is particularly important for distinguishing asystole from coarse ventricular fibrillation.

To restore amplitude information, the base-ten logarithm of the segment standard deviation, measured before normalization, is concatenated with the pooled feature vector. This modification introduces only 64 additional parameters per network. As shown in Section~\ref{sec:ablation}, it increases asystole specificity from 63.5 to 89.3 percent without reducing ventricular-fibrillation sensitivity.

The resulting feature is comparable between the training and test cohorts because both datasets are stored in millivolts and neither is rescaled. Their spectral characteristics are also broadly similar. For sinus rhythm, which carries the strictest specificity requirement, the fraction of spectral energy above 25~Hz is 0.0305 in the training data and 0.0290 in the test data. Supraventricular tachycardia is the only substantial mismatch, with corresponding values of 0.0068 and 0.0269. Nevertheless, specificity for this rhythm remains between 94.40 and 96.06 percent across the twelve test conditions.

\subsection{Training Augmentation}
\label{sec:augmentation}

Every training segment is augmented before being presented to the network. One of the 32 training artifact recordings is selected at random and mixed with the clean ECG at a signal-to-artifact ratio (SAR) drawn uniformly from $[-6,0]$~dB. A clean segment $s$ and an artifact segment $a$ are combined as
\begin{equation}
x = s + \sqrt{\frac{P_s}{10^{\mathrm{SAR}/10}\,P_a}}\; a ,
\label{eq:sar}
\end{equation}
where $P_s$ and $P_a$ denote the powers of the clean ECG and artifact segments, respectively, computed over the window being mixed. In this formulation, the denominator contains the power of the artifact alone. The same definition is used throughout the test conditions.

This SAR formulation is consistent with the studies used for comparison in this work \cite{hajeb2021deep}. Other studies instead define SAR using the power of the mixture in the denominator \cite{jekova2021optimization}. Consequently, SAR values reported in decibels are not always directly comparable across studies.

Clinical noise is added to 70 percent of the training segments. When noise augmentation is applied, electrode-motion artifact is always included, whereas baseline wander and muscle artifact are added with probabilities of 0.8 and 0.4, respectively. Each noise component is mixed using the same formulation as Eq.~(\ref{eq:sar}), with a signal-to-noise ratio drawn uniformly from $[3,20]$~dB.

Three additional augmentation procedures are designed to compensate for rhythm classes that are rare in the training pool, two of which specifically target asystole.

The first generates synthetic asystole because the retained training examples do not cover the amplitude range most relevant to the test cohort. The 45 retained asystole segments have standard deviations ranging from 0.0204 to 0.0788~mV, whereas asystole in the test cohort spans 0.003 to 0.035~mV. To bridge this gap, noise segments from the training pool are mixed and scaled to a target standard deviation drawn log-uniformly from $[0.002,0.05]$. This range was selected to cover the amplitude distribution observed in the deployment cohort and was not adjusted thereafter.

The second augmentation uses a training compression-artifact recording by itself. Because these recordings were collected during asystole, this procedure provides genuine examples of asystole under chest compressions rather than a synthetic approximation.

The third augmentation increases the proportion of rapid supraventricular tachycardia. A narrow-complex non-shockable segment is resampled to a target heart rate drawn uniformly from $[170,220]$ beats per minute, and the resulting waveform is tiled to refill the analysis window. The resampling factor is computed from the measured heart rate of the source segment rather than using a fixed multiplier. For example, a fixed multiplier of 1.8 would map 100 beats per minute to 180, below the median rate of rapid supraventricular tachycardia in the test cohort, while mapping 160 beats per minute to 288, which is not physiologically plausible.

The three augmentation procedures replace a training segment with probabilities of 0.04, 0.06, and 0.06, respectively.

\subsection{Training and Checkpoint Selection}

Each training epoch draws 20{,}000 samples by resampling records to the class ratio associated with the network being trained. Because the contamination is generated anew at every presentation, an oversampled segment reappears with different artifacts, signal-to-artifact ratios, and noise realizations rather than as an exact duplicate.

Optimization uses Adam with a learning rate of $10^{-3}$, weight decay of $10^{-4}$, gradient clipping to unit norm, and a batch size of 256. Training proceeds for up to 300 epochs with early stopping based on validation average precision, and the checkpoint with the highest validation average-precision score is retained.

Average precision was selected because it evaluates ranking performance rather than performance at a single decision threshold, making it an appropriate metric for an imbalanced positive class \cite{saito2015precision}. Alternative criteria, including validation loss, plain accuracy, and balanced accuracy, were evaluated but rejected. These criteria respectively favored very early checkpoints, emphasized the majority class at the natural validation prevalence, or fluctuated with sensitivity and specificity at the fixed threshold of 0.5.

\subsection{Fitting the Operating Point}
\label{sec:threshold}

The operating threshold is determined using the validation data and is subsequently applied unchanged to the test cohort. A default threshold of 0.5 is not appropriate in this setting because the AHA criteria impose different requirements on sensitivity and specificity. In practice, specificity is the more difficult criterion to satisfy. Achieving the required specificity therefore requires an operating threshold above the default value, at the cost of some sensitivity.

The threshold-selection procedure is defined before any test results are examined. Candidate thresholds are first filtered to those whose worst-case validation sensitivity satisfies the AHA requirements at a compression signal-to-artifact ratio of $-3$~dB. Among the remaining candidates, the threshold achieving the highest specificity is selected. This procedure explicitly reflects the asymmetric nature of the AHA criteria by enforcing the required sensitivity floor before optimizing specificity.

Two refinements were required before the threshold-selection procedure became meaningful. First, the test cohort is evaluated on a per-subject basis, whereas validation was initially performed on individual segments. Segment-level aggregation allows recordings contributing many windows to exert disproportionate influence relative to recordings contributing only a few. To align validation with the test protocol, predictions are first averaged within each source recording. This adjustment increased the selected threshold from 0.62 to 0.70 and improved the worst-case validation margin from 0.04 to 0.21 percentage points.

Second, the validation pool contains rhythm classes that are absent in the test cohort. The specificity constraint is therefore computed only over rhythm categories present in both datasets, ensuring that threshold selection is aligned with the population on which performance will ultimately be evaluated.

A single operating threshold is fitted using all 15 train-validation splits rather than fitting a separate threshold for each split. This choice reflects the intended deployment scenario, in which a device operates with a single fixed operating point. The same threshold is therefore applied to all three networks.

Threshold selection is performed separately for each model by exhaustive search over a grid with a spacing of 0.04, yielding a threshold of 0.70. An alternative strategy using independently optimized thresholds for the three networks was also evaluated. Although this approach increases the minimum validation margin by 0.05 percentage points, it reduces coverage from 98.4 to 91.7 percent. The shared threshold is therefore retained because it provides a better balance between margin and coverage.

%% file: 05_experimental_setup_R2.tex
\section{Experimental Setup}
\label{sec:setup}

\subsection{Test Conditions}

Each test condition is constructed by adding compression artifact to a clean recording from the Defibtech cohort. Two artifact levels are considered, 0~dB and $-3$~dB, and each is combined with six noise scenarios, giving twelve test conditions. The six scenarios are: no added noise, electrode motion alone, baseline wander alone, muscle artifact alone, electrode motion with baseline wander, and all three noise sources together. Whenever noise is present, it is added at 10~dB. Every subject is evaluated with each of the four test artifacts of that split. As a result, each condition contains 1{,}856 samples and is evaluated across all 15 train-validation splits.

Two conditions are referred to repeatedly throughout this work. The \emph{fitting condition} consists of compression artifact at $-3$~dB with no added noise. It is the most challenging artifact level in the evaluation grid and is used for operating-threshold selection. The \emph{headline condition} consists of compression artifact at 0~dB with no added noise. Summary performance values are reported under this condition. Because a rate measured in one condition does not necessarily predict the rate in the other, every reported result identifies the condition under which it was obtained.

\subsection{Metric Definitions}
\label{sec:metrics}

All performance measures are computed per subject and then averaged within each rhythm group. This prevents subjects with multiple test samples from exerting disproportionate influence on the reported rates. Because each subject appears once for each of the four test artifacts, pooling samples directly would count the same individual four times. Confidence intervals are bootstrap percentile intervals computed over subjects with 2{,}000 resamples \cite{efron1993bootstrap}.

Two conventions are available for scoring an indeterminate outcome. The primary convention used in this work is clinical intention-to-diagnose (ITD), under which an indeterminate outcome is scored as a non-shockable classification. Accordingly, it is counted as an error for a shockable rhythm and as a correct classification for a non-shockable rhythm. The indeterminate state exists only within the cascade, which emits one of two
labels for every segment. ITD therefore records what the classifier returns
rather than excluding difficult cases from evaluation. It also places methods with and without abstention mechanisms on the same footing. Under the excluded-indeterminate (EI) convention, indeterminate outcomes are removed from the denominator and their frequency is reported separately. This typically increases sensitivity and decreases specificity. EI is used by much of the existing literature and is therefore reported alongside the published values in Table~\ref{tab:comparison}.

\subsection{Performance Criteria}
\label{sec:criteria}

The AHA recommendations define performance requirements for five rhythm categories, each accompanied by a minimum test sample size \cite{kerber1997automatic}. IEC 60601-2-4 adopts the same performance targets \cite{iec60601_2_4}. Sensitivity must exceed 90 percent for coarse ventricular fibrillation, evaluated on at least 200 episodes, and 75 percent for rapid ventricular tachycardia, evaluated on at least 50 episodes. Specificity must exceed 99 percent for sinus rhythm on at least 100 episodes, and 95 percent for both asystole and the other non-shockable rhythms, evaluated on at least 100 and 30 episodes, respectively. The recommendations treat asystole as a separate category rather than as part of the non-shockable group.

The AHA criteria were developed for recordings free of compression artifact, because the devices they describe analyze the rhythm during pauses in chest compressions. Every condition considered in this work therefore evaluates a more difficult problem than the one for which the criteria were written. The same numerical thresholds are applied because no alternative performance standard exists for rhythm analysis during ongoing compressions. Compliance in this work should therefore be interpreted as compliance with the AHA criteria under compression artifact, a stricter test than the recommendations require for a device that analyzes during compression pauses.

The test cohort meets the recommended minimum sample size in only two of the five AHA categories. Sinus rhythm contributes 169 subjects against a minimum requirement of 100, and the other non-shockable category contributes 214 against 30. Coarse ventricular fibrillation contributes 41 subjects against 200, rapid ventricular tachycardia 31 against 50, and asystole 9 against 100. Compliance is therefore evaluated only for sinus rhythm and the other non-shockable rhythms. Results for the remaining categories are reported as measurements rather than compliance assessments. Asystole is reported because it is the rhythm class this work specifically aims to improve, not because nine subjects are sufficient to establish a reliable level of performance.

\subsection{Comparisons}

Three comparisons are reported. 
The first comparison tests whether the gains of the proposed method arise from its routing strategy rather than from model capacity. The cascade is replaced by two single-network baselines with matched parameter counts. The first is a widened version of the proposed architecture, obtained by scaling every channel dimension by 1.74 to give channel counts of 56, 111, 223, and 223 and a total of 1{,}323{,}755 parameters, and trained as a balanced classifier with no routing and no abstention. The second is a dilated temporal convolutional network (TCN) \cite{bai2018empirical} with 142 channels, seven dilation levels, and a kernel size of 5, giving 1{,}326{,}899 parameters and a receptive field of 1{,}017 samples that spans the full eight-second window. Both baselines use the same train-validation splits, augmentations, and threshold-selection procedure as the cascade. The architecture itself is therefore the only experimental variable.

The second comparison examines the abstention strategy independently of coverage. Four alternative abstention rules are constructed from networks already trained, so none adds parameters. Three are derived from the balanced network alone and the fourth from the majority decision of the three networks. Each rule is constrained to abstain on exactly as many segments as the cascade. Without matching coverage, a comparison could not distinguish a more effective abstention strategy from one that simply abstains more often \cite{geifman2017selective}. Fixing the abstention rate therefore isolates the quality of the decision rule itself. The four rules abstain, respectively, on the segments whose predicted probability lies closest to the operating threshold, on those with the smallest margin between the two class scores, on those with the highest binary entropy, and on those for which the three networks reach no majority decision. For a single sigmoid output, margin and binary entropy order the segments identically and therefore select the same segments.

The third comparison separates the contribution of the amplitude input from that of the augmentation strategy. Each is removed in turn while the other is retained. The resulting configuration quantifies the benefit of the augmentations without conflating it with the gain provided by the amplitude input.

\subsection{Implementation}
\label{sec:imp}
Inference latency is measured on a single thread of an AMD EPYC 7H12 processor with one segment presented at a time, and reported as the median of 300 timed repetitions with an interquartile range. This reflects the intended deployment setting, where a defibrillator analyzes one window at a time on a limited processor. Batch throughput on a data-center accelerator is therefore not used as a deployment metric. The cascade is timed along the path a decision takes, which evaluates the balanced network and one biased network rather than all three. Models were trained and evaluated in PyTorch 2.5.1 under Python 3.12.5; latency and quantization were measured in PyTorch 2.13.0 under Python 3.14.6 after a cluster upgrade retired the earlier runtime.

%% file: 06_results_R2.tex
\section{Results}
\label{sec:results}

Unless otherwise stated, all results in this section are reported under the clinical ITD convention and averaged across the 15 train-validation splits. The reported values correspond to the default configuration: an eight-second analysis window and a decision threshold of 0.70 selected on the validation data. A fourteen-second window was evaluated in the same manner and satisfies the adjudicated criteria under the same six conditions. However, the eight-second window is reported throughout because it achieves the same outcome with a shorter analysis interval.

\subsection{Performance Across the Test Conditions}

Table~\ref{tab:main} summarizes performance across the six conditions at 0~dB, including the five AHA rates, the indeterminate share, and the compliance verdict where adjudication is possible. The six conditions at $-3$~dB are reported in the text below.

\begin{table*}[!t]
\caption{Performance across the six test conditions at a compression signal-to-artifact ratio of 0~dB using an eight-second analysis window and a decision threshold of 0.70. The corresponding results at $-3$~dB are described in the text. Subscripts indicate the standard deviation across the 15 train-validation splits, which differ in both the training partition and the four compression artifacts used at test time. For the sample-size reasons outlined in Section~\ref{sec:criteria}, AHA compliance is adjudicated only for sinus rhythm and the other non-shockable rhythms. $\SPO$ is computed over the 214 non-shockable subjects excluding asystole, consistent with the AHA definition of the category, whereas $\SPOA$ additionally includes the 9 asystole subjects. All values are reported as percentages.}
\label{tab:main}
\centering
\small
\begin{tabular}{@{}lrrrrrrrc@{}}
\toprule
 Added noise & \SEVF & \SEVT & \SPN & \SPA & \SPO & \SPOA &
Indet. & AHA compliance \\
\midrule
 none           & $98.21_{3.4}$ & $92.69_{4.7}$ & $99.36_{0.6}$ & $86.67_{10.9}$ & $95.92_{2.0}$ & $95.55_{2.3}$ & 1.19 & Pass \\
 EM             & $98.29_{2.6}$ & $93.12_{4.5}$ & $99.20_{0.6}$ & $87.04_{11.9}$ & $95.48_{2.3}$ & $95.14_{2.6}$ & 1.18 & Pass \\
 BW             & $97.85_{3.4}$ & $92.26_{5.3}$ & $99.34_{0.6}$ & $87.59_{11.2}$ & $95.97_{2.0}$ & $95.64_{2.2}$ & 1.07 & Pass \\
 MA             & $97.11_{4.3}$ & $91.34_{6.8}$ & $99.43_{0.6}$ & $89.26_{11.0}$ & $96.20_{2.2}$ & $95.92_{2.4}$ & 1.22 & Pass \\
 EM + BW        & $97.72_{2.7}$ & $92.20_{4.7}$ & $99.22_{0.7}$ & $88.70_{10.3}$ & $95.72_{2.1}$ & $95.43_{2.4}$ & 1.26 & Pass \\
 EM + BW + MA   & $95.77_{5.6}$ & $90.27_{8.7}$ & $99.21_{0.9}$ & $91.48_{10.2}$ & $95.87_{2.4}$ & $95.70_{2.6}$ & 1.48 & Pass \\
\bottomrule
\end{tabular}
\end{table*}

Across all six conditions at 0~dB, the cascade satisfies both adjudicable AHA criteria. Sinus-rhythm specificity ranges from 99.20 to 99.43 percent, exceeding the required 99 percent on a cohort of 169 subjects. Specificity for the other non-shockable rhythms ranges from 95.48 to 96.20 percent, exceeding the required 95 percent on 214 subjects. The indeterminate rate remains low at 1.07 to 1.48 percent, meaning that more than 98.5 percent of segments receive a definitive shockable or non-shockable classification.

The remaining three rates are reported as descriptive measurements rather than compliance assessments because the cohort does not meet the recommended minimum sample size for those categories (Section~\ref{sec:criteria}). Sensitivity ranges from 95.77 to 98.29 percent for coarse ventricular fibrillation and from 90.27 to 93.12 percent for ventricular tachycardia, both exceeding their respective AHA criteria. Asystole specificity ranges from 86.67 to 91.48 percent, below the nominal 95 percent target. However, the available sample size is insufficient to support a formal compliance determination for any of these three categories.

The subscripts in Table~\ref{tab:main} quantify variation across the 15 train-validation splits, which differ in both the training partition and the four artifacts used at test time. Sampling uncertainty is a separate quantity and is estimated using bootstrap percentile intervals computed from 2{,}000 resamples of the subject cohort at the headline condition. These intervals support the same conclusions as the mean values. For the two adjudicated categories, the lower bounds remain above the corresponding AHA criteria, reaching 99.09 percent for sinus rhythm and 95.34 percent for the other non-shockable rhythms. In contrast, asystole remains below its nominal target across the entire interval, which spans 81.66 to 91.11 percent. The most challenging noise condition, combining all three noise sources, reduces ventricular-fibrillation sensitivity by 2.4 percentage points and sinus-rhythm specificity by 0.15 percentage points. Even under this condition, both adjudicated AHA criteria remain satisfied.

At $-3$~dB, performance degradation is driven primarily by loss of specificity rather than sensitivity. As a result, none of the six conditions satisfies the two adjudicated AHA criteria. Sinus-rhythm specificity falls to between 98.43 and 98.67 percent, remaining 0.33 to 0.57 percentage points below the required 99 percent. Specificity for the other non-shockable rhythms falls to between 93.89 and 94.54 percent, remaining 0.46 to 1.11 percentage points below the required 95 percent. In contrast, sensitivity remains well above the corresponding AHA requirements for both shockable rhythms in every condition. Ventricular-fibrillation sensitivity never falls below 92.07 percent, and ventricular-tachycardia sensitivity never falls below 89.84 percent. Asystole specificity declines further, ranging from 85.00 to 90.00 percent.

\subsection{Error Analysis by Rhythm}

\begin{figure}[!t]
\centering
\includegraphics[width=\figRhythmW]{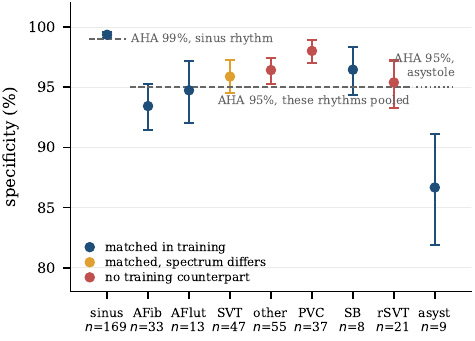}
\caption{Rhythm-specific specificity at the headline condition (compression artifact at 0~dB with no added noise), evaluated under the clinical ITD convention. Error bars show 95 percent bootstrap percentile intervals over subjects, and the number of subjects is shown beneath each rhythm label. Rhythms are grouped according to whether a corresponding class is present in the training pool (Table~\ref{tab:testcohort}). Supraventricular tachycardia is shown separately because its training counterpart differs substantially in spectral characteristics. The applicable AHA specificity criteria are described in Section~\ref{sec:criteria}.}
\label{fig:rhythm}
\end{figure}

Fig.~\ref{fig:rhythm} summarizes rhythm-specific specificity at the headline condition. To facilitate interpretation, the rhythms are grouped according to whether the training pool contains a corresponding class.

Specificity exceeds 91 percent for every rhythm except asystole. The width of the bootstrap intervals largely reflects sample size, ranging from approximately two percentage points for sinus rhythm (169 subjects) to about twenty percentage points for asystole (9 subjects). Importantly, the relevant AHA criterion applies to the non-shockable rhythms other than sinus rhythm and asystole as a combined category. Consequently, a single rhythm falling below the 95 percent threshold does not by itself imply non-compliance.

Rapid supraventricular tachycardia merits separate discussion because it is among the least frequent rhythms in the training data. The class has no direct training counterpart, and its median heart rate of 197~beats per minute exceeds that of nearly every segment in the training pool. Only 96 non-shockable training segments exceed 160~beats per minute. Consequently, the network encounters this rate range primarily through the time-compression augmentation described in Section~\ref{sec:augmentation}. Despite this limited training exposure, specificity remains between 95.40 and 96.27 percent across the six conditions at 0~dB.

\subsection{Comparison with Matched-Parameter Baselines}
\label{sec:baselines}

Table~\ref{tab:baselines} compares the proposed cascade with two matched-parameter single-network baselines. To ensure a fair comparison, all three models use thresholds determined by the same validation-based procedure on the same validation records and under the same fitting condition. Each model is evaluated using its own validation-derived threshold rather than a threshold selected from another architecture or from the test cohort.

\begin{table}[!t]
\caption{Comparison with matched-parameter single-network baselines at the eight-second window. Reported values correspond to the worst-performing condition among the twelve evaluated test conditions, with subscripts indicating the standard deviation across the 15 train-validation splits. Because the baseline models do not include an abstention mechanism, the ITD and EI conventions produce identical rates for both baselines. The compliance column reports the number of the six 0~dB conditions that satisfy the adjudicated AHA criteria. For the proposed cascade, the reported latency corresponds to a single decision, which evaluates the balanced network and the selected biased network.}
\label{tab:baselines}
\centering
\setlength{\tabcolsep}{2pt}
\scriptsize
\begin{tabular}{@{}lrrrcccc@{}}
\toprule
Model & Params & Thr. & Time & Worst-case & Worst-case & Worst-case & Compliance \\
      &        &      & (ms) & \SEVF      & \SPN & \SPO & at 0~dB \\
\midrule
E2E cascade & 1{,}325{,}763 & 0.70 &  6.68 & $92.07_{10.4}$ & $98.43_{1.6}$ & $93.89_{3.8}$ & 6 / 6 \\
Wide CNN    & 1{,}323{,}755 & 0.62 &  6.72 & $94.84_{13.8}$ & $94.80_{10.3}$ & $87.63_{13.0}$ & 0 / 6 \\
TCN         & 1{,}326{,}899 & 0.88 & 39.67 & $91.67_{11.5}$ & $98.45_{1.4}$ & $93.06_{4.0}$ & 5 / 6 \\
\bottomrule
\end{tabular}
\end{table}

The clearest distinction among the three models is observed in sinus-rhythm specificity. Both single-network baselines fall short of the 99 percent AHA criterion, whereas the proposed cascade satisfies it under all six adjudicated 0~dB conditions. Across the twelve test conditions, the widened network achieves a worst-case sinus-rhythm specificity of 94.80 percent. The corresponding worst-case values are 98.45 percent for the dilated network and 98.43 percent for the cascade. Under the six 0~dB conditions used for compliance assessment, the widened network reaches a minimum specificity of 96.83 percent, while the dilated network reaches 98.97 percent and satisfies the criterion in five of the six conditions. The cascade reaches 99.20 percent and satisfies the criterion in all six conditions.

Sensitivity alone does not distinguish the three models. At the headline condition, the widened network achieves 97.76 percent ventricular-fibrillation sensitivity, close to the 98.21 percent achieved by the cascade. The primary difference emerges in specificity, where the widened network sacrifices enough performance to fail the sinus-rhythm criterion.

Inference cost likewise does not account for the observed advantage. Although the cascade contains three networks, each decision evaluates only two of them: the balanced network and the biased network selected by the routing rule. Measured along this decision path in the same benchmark, the cascade requires 6.68~ms per decision compared with 6.72~ms for the widened network, a difference smaller than the interquartile range of either measurement. The two models are therefore effectively matched in both parameter count and inference cost, indicating that the observed improvement is primarily attributable to the routing strategy.

Differences between models are evaluated using a paired Wilcoxon signed-rank test across the 15 train-validation splits. Pairing is performed by split, ensuring that each comparison uses identical training and validation partitions. Compared with the widened network, the cascade achieves significantly higher sinus-rhythm specificity in all twelve conditions after Holm correction, with gains ranging from 1.85 to 3.63 percentage points. In contrast, no statistically significant differences are observed between the cascade and the dilated network for any of the five reported performance metrics under any test condition.

The absence of statistical significance should be interpreted cautiously. Although the differences do not reach statistical significance, the direction of the effect is highly consistent: the cascade outperforms the dilated network in 56 of 60 paired comparisons. This pattern is more consistent with limited statistical power than with true equivalence between the two models.

Asystole provides the clearest example. Depending on the test condition, the cascade achieves 1.67 to 6.30 percentage points higher specificity than the dilated network. However, the class contains only nine subjects, and the standard deviation across the 15 train-validation splits ranges from 10.2 to 23.5 percentage points across the two models. Under variability of this magnitude, performance differences of only a few percentage points are difficult for a significance test to resolve reliably.

Differences in variability across the train-validation splits distinguish the models more clearly than differences in their mean performance. Under the worst-performing condition, the standard deviation of sinus-rhythm specificity is 1.6 percentage points for the cascade and 1.4 percentage points for the dilated network, compared with 10.3 percentage points for the widened network. The statistical test therefore identifies the instability of the widened model much more readily than the relatively small performance differences between the cascade and the dilated network.

Consequently, the present analysis does not provide statistical evidence that the cascade outperforms the dilated network. Their practical differences emerge elsewhere. The dilated network requires 39.67~ms per decision compared with 6.68~ms for the cascade, satisfies five rather than all six adjudicated 0~dB compliance conditions, and always produces a classification rather than providing an abstention mechanism.

\subsection{Comparison of Abstention Strategies at Matched Coverage}
\label{sec:abstention}

An abstention mechanism can always reduce the observed error rate simply by declining to classify more difficult segments. The key question, therefore, is not whether a method abstains, but whether it abstains on the right segments. Addressing that question requires coverage to be matched across methods.

Table~\ref{tab:abstention} compares the proposed cascade with four alternative abstention strategies under matched coverage. Each alternative is constrained to abstain on exactly the same number of segments as the cascade, after which the error rate is computed over the segments that remain classified. Matching coverage ensures that observed differences reflect the quality of the abstention rule itself rather than differences in how often a method abstains.
\begin{table}[!t]
\caption{Comparison of abstention strategies under matched coverage. Reported values are the error rate (percent) among the segments that receive a classification, with lower values indicating more effective identification of difficult cases. Because abstained segments are excluded from the calculation, the ITD and EI scoring conventions yield identical results for this comparison.}
\label{tab:abstention}
\centering
\setlength{\tabcolsep}{3pt}
\footnotesize
\begin{tabular}{@{}lrrrrr@{}}
\toprule
Condition & Cascade & Confidence & Margin & Entropy & Vote \\
\midrule
0~dB none        & \textbf{2.817} & 2.941 & 3.304 & 3.304 & 3.202 \\
0~dB EM         & \textbf{3.086} & 3.170 & 3.518 & 3.518 & 3.344 \\
0~dB BW          & \textbf{2.839} & 2.992 & 3.333 & 3.333 & 3.134 \\
0~dB MA          & \textbf{2.698} & 2.953 & 3.233 & 3.233 & 3.193 \\
0~dB EM+BW      & \textbf{2.979} & 3.165 & 3.489 & 3.489 & 3.427 \\
0~dB all three   & \textbf{2.986} & 3.303 & 3.646 & 3.646 & 3.351 \\
$-3$~dB none     & \textbf{4.056} & 4.261 & 4.802 & 4.802 & 4.404 \\
$-3$~dB EM      & \textbf{4.326} & 4.618 & 5.105 & 5.105 & 4.710 \\
$-3$~dB BW       & \textbf{4.175} & 4.339 & 4.895 & 4.895 & 4.522 \\
$-3$~dB MA       & \textbf{4.130} & 4.533 & 4.965 & 4.965 & 4.939 \\
$-3$~dB EM+BW   & \textbf{4.383} & 4.719 & 5.136 & 5.136 & 4.971 \\
$-3$~dB all three& \textbf{4.348} & 4.821 & 5.282 & 5.282 & 4.887 \\
\bottomrule
\end{tabular}
\end{table}

The cascade achieves the lowest answered-segment error rate in all twelve test conditions. Relative to the strongest competing rule in each condition, the improvement ranges from 2.7 to 10.9 percent when expressed as a fraction of the cascade's own error rate. Relative to the weakest rule, the improvement reaches 22.1 percent.

Among the alternatives, the confidence rule is consistently the strongest competitor. By abstaining on the segments whose predicted probabilities lie closest to the decision threshold, it is the most competitive single-network strategy. Nevertheless, the cascade achieves lower error rates in every condition. These results indicate that disagreement between specialized networks identifies difficult segments more effectively than confidence alone. The advantage becomes larger as the interference conditions become more severe.

Differences between abstention strategies are evaluated using paired Wilcoxon signed-rank tests across the twelve test conditions. In every condition, the cascade achieves a lower answered-segment error rate than each alternative strategy. Averaged across the twelve conditions, the advantage is 0.249 percentage points relative to the confidence rule, 0.438 percentage points relative to the vote rule, and 0.657 percentage points relative to the margin and entropy rules, with $p = 0.0005$ in all three comparisons.

Because all twelve conditions are evaluated using the same cohort and artifact set, they should not be interpreted as twelve independent experiments. Instead, the statistical analysis demonstrates the consistency of the observed advantage across conditions rather than twelve independent confirmations of the effect.

\subsection{Augmentation Ablation}
\label{sec:ablation}

Data augmentation plays a critical role in the performance of the proposed cascade. To quantify its contribution, the model is retrained without the three augmentation procedures while retaining the amplitude input and the same 15 train-validation splits. Removing these augmentations eliminates compliance in all six adjudicated 0~dB conditions, compared with six of six conditions when they are included. Most of the benefit arises from improved classification of asystole and other non-shockable rhythms. Asystole specificity increases by 12 to 16 percentage points, while specificity for the other non-shockable rhythms increases by 1.2 to 1.4 percentage points. In contrast, ventricular-fibrillation sensitivity changes little, increasing by 0.3 percentage points at 0~dB and decreasing by 0.4 percentage points at $-3$~dB.

\begin{figure}[!t]
\centering
\includegraphics[width=\figAmplitudeW]{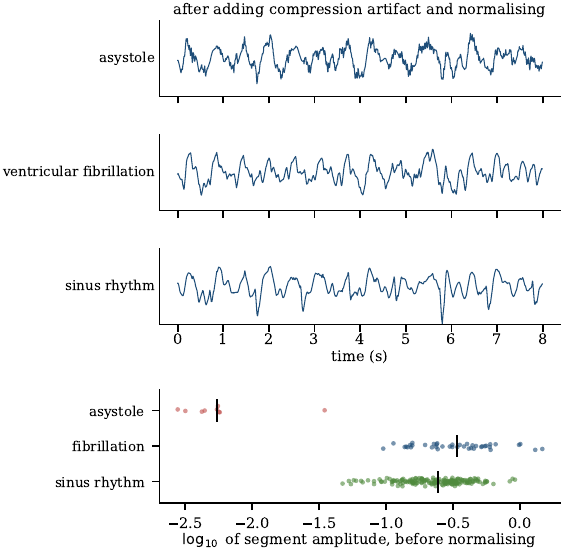}
\caption{Illustration of the motivation for the amplitude input. Top: representative asystole, ventricular-fibrillation, and sinus-rhythm recordings from the test cohort after the addition of compression artifact at 0~dB and subsequent normalization, matching the network input. Bottom: distribution of segment standard deviation before normalization. After normalization, asystole and ventricular-fibrillation signals become visually similar despite their substantial amplitude differences. Because normalization removes this diagnostic amplitude information, distinguishing the two rhythms becomes more difficult unless amplitude is provided as a separate input.}
\label{fig:amplitude}
\end{figure}

The contribution of the amplitude input is evaluated separately by retraining the fully augmented cascade without it on the same 15 train-validation splits. At the headline condition, including the amplitude input increases asystole specificity from 63.5 to 89.3 percent and reduces the standard deviation across splits from 22.7 to 7.0 percentage points, restoring diagnostic amplitude information that would otherwise be removed by normalization. Ventricular-fibrillation sensitivity remains essentially unchanged. Because the ablation experiments are independent training runs, their absolute values differ slightly from those reported in Table~\ref{tab:main} due to run-to-run variation, which is greatest for asystole. The effect is illustrated in Fig.~\ref{fig:amplitude}.

%% file: 06a_deployment_R2.tex
\section{Deployment Considerations}
\label{sec:deployment}

The second requirement stated in Section~\ref{sec:intro} is that a decision be reached on the processor inside the AED. On-device inference of this kind has become increasingly important in resource-constrained connected devices, where local execution reduces latency and dependence on remote computation \cite{jhaveri2024tinyml}, and neural networks for ventricular-fibrillation detection have been explored on ARM-based microcontrollers \cite{krasteva2025implementing}. This section characterizes the arithmetic, memory footprint, and latency of the cascade, together with the effect of eight-bit quantization, so that its requirements can be compared with the resources of such processors.

Although the cascade contains three networks, each decision evaluates only two of them, the balanced network and the selected biased network, as indicated in Fig.~\ref{fig:system}(d). Consequently, an analysis requires two forward passes rather than three. Each network performs 51.5 million multiply-accumulate operations on an eight-second window, resulting in approximately 103 million operations per decision. All three networks must nevertheless remain in memory because either biased network may be selected at run time. In single precision, this corresponds to 1.33 million parameters and 5.15~MiB of model state, including batch-normalization statistics.

Latency was measured with one segment presented at a time on a single thread of an AMD EPYC 7H12 processor using the protocol described in Section~\ref{sec:imp}. A single decision requires 6.68~ms, with an interquartile range of 6.65 to 6.71~ms across 300 timed repetitions. Evaluating all three networks, as performed during offline analysis to sweep operating points without repeating inference, requires 9.96~ms.

The measured latency leaves a wide margin relative to the analysis interval. The same decision would complete within one second on a processor approximately 150 times slower than the one used here, and on the present processor it occupies only 0.08 percent of an eight-second window. Even if rhythm analysis were repeated once per second, the corresponding duty cycle would remain 0.67 percent. These margins indicate that classifier execution is unlikely to be the primary factor governing processor selection. The figures nevertheless characterize a general-purpose processor, and a uniformly slower processor is only a coarse proxy for a microcontroller, whose throughput also depends on its memory hierarchy, arithmetic support, and inference runtime. Deep networks for ventricular-fibrillation detection have been implemented on ARM-based microcontrollers \cite{krasteva2025implementing}, indicating that networks for this task can execute on such hardware, but the present cascade has not been measured on a target device. Table~\ref{tab:deployment} summarizes the resulting deployment metrics.

\begin{table}[!t]
\centering
\caption{Computational and memory requirements of the proposed cascade at the eight-second analysis window. Latency is reported as the median of 300 timed repetitions with one segment presented at a time on a single thread of an AMD EPYC 7H12 processor. Model state includes all three networks and their batch-normalization statistics, since the cascade retains all networks in memory and selects between the two biased networks at run time.}
\label{tab:deployment}
\footnotesize
\begin{tabular}{@{}lr@{}}
\toprule
Quantity & Value \\
\midrule
Parameters, three networks        & 1{,}325{,}763 \\
Multiply-accumulates per decision & $103\times10^{6}$ \\
Model state, single precision         & 5.15~MiB \\
Model state, eight-bit                & 1.45~MiB \\
Latency per decision              & 6.68~ms \\
Duty cycle, one decision per window & 0.08~\% \\
\bottomrule
\end{tabular}
\end{table}

Weight storage was evaluated using eight-bit post-training quantization with static calibration on the validation data. The convolutional layers were quantized, whereas the pooling operation, the concatenation with the log-amplitude feature, and the classifier head were retained in single precision. This design choice was motivated by the differing numerical ranges of the pooled feature vector and the log-amplitude input. The pooled features are non-negative, whereas the log amplitude is negative for all rhythm categories and can reach approximately $-2.5$ for asystole. A single affine quantization scale spanning both quantities therefore uses its available resolution inefficiently.

Quantizing the classifier head as well reduces model storage by an additional 4.3 percent, but increases the largest change in predicted probability by more than two orders of magnitude. Given that the classifier head contains only 8{,}385 of the 441{,}921 parameters in a network, this trade-off was not considered favorable. Retaining the classifier head in single precision reduces the storage requirement of the cascade to 1.45~MiB, corresponding to a 3.55-fold reduction relative to the single-precision model.

Because the processor used for timing does not provide integer dot-product acceleration, quantization is evaluated here primarily in terms of storage and classification performance rather than inference speed.

Using the operating threshold determined for the single-precision cascade, quantization had minimal impact on classification performance. Across the twelve test conditions, sinus-rhythm specificity changed by at most 0.07 percentage points and specificity for the other non-shockable rhythms by at most 0.08 percentage points. These are the two rhythm categories for which the cohort is sufficiently large to support AHA adjudication.

Ventricular-fibrillation and ventricular-tachycardia sensitivities changed by at most 0.16 and 0.33 percentage points, respectively. Asystole specificity changed by at most 1.48 percentage points, corresponding to fewer than one effective classification difference across the evaluated cohort. Most importantly, every compliance verdict remained unchanged across all twelve test conditions under both the four-category and five-category interpretations of the criteria.

Refitting the operating threshold on validation data using the same procedure applied to the single-precision cascade selected a threshold of 0.66 rather than 0.70. After quantization, the limiting validation criterion shifted from ventricular-fibrillation sensitivity to sinus-rhythm specificity, and the operating margin at the selected threshold decreased from 0.21 to 0.15 percentage points. These changes are small but indicate that quantization can alter the optimal operating point. Consequently, a quantized classifier should have its threshold refitted rather than simply inheriting the threshold selected for the single-precision model.

%% file: 07_discussion_R2.tex
\section{Discussion}
\label{sec:discussion}

\subsection{What the Cascade Contributes}

The AHA performance requirements are inherently asymmetric. Specificity for sinus rhythm must exceed 99 percent, whereas sensitivity for coarse ventricular fibrillation must exceed only 90 percent. A classifier trained to minimize a single undifferentiated error is not explicitly optimized for such an operating point. The central question is therefore what the proposed cascade contributes beyond simply using multiple networks. The results of Section~\ref{sec:results} allow three possible explanations to be examined separately. The observed advantage could arise from increased model capacity, from the underlying network architecture, or from abstaining on difficult segments. The results provide little evidence that model capacity or abstention alone explains the improvement. Instead, they indicate that the routing strategy is the primary contributor, although the comparison with the dilated network remains a trade-off rather than a definitive advantage.

The remaining explanation is the routing mechanism itself. The cascade does not simply combine multiple classifiers; it assigns them distinct roles. A shockable classification is verified by a network trained to favor shockable outcomes, whereas a non-shockable classification is verified by a network trained to favor non-shockable outcomes. Disagreement between the initial and verification stages therefore carries a specific interpretation. It indicates that the evidence is weak even when evaluated by the classifier most inclined toward the proposed outcome. This information is not captured solely by confidence scores or proximity to a decision threshold. The results indicate that exploiting this disagreement allows the cascade to identify uncertain segments more effectively than the alternative abstention strategies evaluated in this study.

\subsection{Comparison With Reported Results}
\label{sec:comparison}

Table~\ref{tab:comparison} summarizes previously reported methods for rhythm analysis during chest compressions alongside the proposed cascade. The purpose of the table is to place the results in context rather than to establish a strict ranking of methods. Direct numerical comparison is difficult because the studies differ in their test populations, compression protocols, analysis windows, train-test separation procedures, and treatment of indeterminate outcomes. These differences are reported explicitly alongside the performance figures. The accompanying methodological details are often as informative as the rates themselves and are required for a fair interpretation of the published results.

\begin{table*}[!t]
\centering
\caption{Published methods for rhythm analysis during chest compressions together with the methodological factors needed for their interpretation$^{\dagger,a}$. ``Aux.'' indicates whether the method requires measurements beyond the electrocardiogram for rhythm analysis. ``Separation'' describes how training and test data are kept independent. ``Indet.'' reports the percentage of analyses that received no confident classification, together with the convention used to calculate the reported performance rates$^{\dagger}$. Ranges indicate variation across the conditions evaluated in the original study. ``n/r'' indicates values that were not reported.}
\label{tab:comparison}
\footnotesize
\resizebox{\textwidth}{!}{%
\begin{tabular}{@{}llcllcrrrrrr@{}}
\toprule
Study & Approach & Aux. & Test data & Separation & Window &
\SEVF & \SEVT & \SPN & \SPO & Indet. & Latency \\
\midrule
\multicolumn{12}{@{}l}{\textit{Adaptive filtering, auxiliary reference signal required}} \\
Irusta 2009 \cite{irusta2009least} & LMS filter, compression rate & Yes &
  OHCA & patient & 14.4~s & 95.6 & n/r & 86.4 & n/r & 0, none & n/r \\
Ruiz 2010 \cite{ruiz2010cardiopulmonary} & Kalman filter, compression rate & Yes &
  OHCA & patient & 14.4~s & 93.3 & n/r & 89.1 & n/r & 0, none & n/r \\
\midrule
\multicolumn{12}{@{}l}{\textit{End-to-end classification, electrocardiogram only}} \\
Hajeb-M 2021 \cite{hajeb2021deep} & Deep network & No &
  Human + CPR & n/r & n/r & 95.2 & n/r & n/r & 86.0 & 0, none & 150~ms \\
Krasteva 2023 \cite{krasteva2023deep} & Sliding CNN & No &
  OHCA & n/r & 10~s & 92--94.4 & n/r & 98.2--99.5 & 96--97 & 0, none & n/r \\
\midrule
\multicolumn{12}{@{}l}{\textit{Artifact removal followed by classification}} \\
Isasi 2019 \cite{isasi2019machine} & RLS filter, SVM & Yes &
  Mechanical CPR & patient & 8~s & 97.5$^{b}$ & n/r & n/r & 98.2$^{b}$ & 0, none & 21~ms \\
Nejad 2024 \cite{nejad2024enhancing} & Cascade CNN encoder-decoder$^{c}$ & No &
  Human + CPR & none$^{c}$ & 14~s & 95.4 & 87.7 & 99.4 & 97.2 & 3.4, EI & 1.1~s \\
Lim 2026 \cite{lim2026cascade} & Cascade autoencoder$^{d}$ & No &
  Defibtech, 464 & artifact & 14~s & 97.5 & 94.6 & 99.3 & 97.2 & 3.6, EI & n/r \\
\midrule
\multicolumn{12}{@{}l}{\textit{Commercial systems evaluated on clinical recordings}} \\
cprINSIGHT 2021 \cite{degraaf2021analyzing} & Proprietary, impedance & Yes &
  OHCA & n/r & n/r & 96 & n/r & 98 & n/r & 30, EI & n/r \\
AWC 2024 \cite{didon2024clinical} & Proprietary & No$^{e}$ &
  OHCA & patient & 16~s & 94.9 & n/r & 100 & 99.3 & 0, none & n/r \\
\midrule
\multicolumn{12}{@{}l}{\textbf{This work (no artifact removal)}} \\
Cascade, 8~s  & Verification cascade & \textbf{No} & Defibtech, 464 & \textbf{patient} &
  \textbf{8~s} & 98.2 & 92.7 & 99.4 & 95.9 & \textbf{1.2, ITD} & \textbf{6.68~ms} \\
Cascade, 8~s, all noise & Verification cascade & \textbf{No} & Defibtech, 464 &
  \textbf{patient} & \textbf{8~s} & 95.8 & 90.3 & 99.2 & 95.9 & 1.5, ITD & \textbf{6.68~ms} \\
\bottomrule
\end{tabular}}
\vspace{1mm}

\raggedright\scriptsize
$^{\dagger}$EI removes indeterminate outcomes from the denominator. Under ITD, an indeterminate outcome is scored as a non-shockable classification, making it an error for a shockable rhythm and a correct classification for a non-shockable rhythm. This convention places methods with and without abstention mechanisms on the same basis. ``none'' indicates that a method has no abstention mechanism. The specificity column for this work is computed over the non-shockable rhythms other than sinus rhythm and asystole, consistent with the AHA definition of the category.
\quad
$^{a}$Latency values are reported as published in each study and were measured on different hardware platforms. They should therefore be interpreted as indicative rather than as a controlled comparison.
\quad
$^{b}$Reported as pooled shockable and non-shockable rates rather than rhythm-specific metrics; shown in the nearest corresponding column.
\quad
$^{c}$All 52 compression artifacts used for both training and testing.
\quad
$^{d}$Ten-fold cross-validation with rotating compression artifacts; the electrocardiogram partition remains fixed.
\quad
$^{e}$Rhythm analysis uses only the electrocardiogram; impedance is used only to confirm the presence of chest compressions.
\end{table*}

An important difference among published studies is how indeterminate outcomes are handled when performance is reported. Every prior method in Table~\ref{tab:comparison} that reports an indeterminate category evaluates sensitivity and specificity under the excluded-indeterminate (EI) convention, whereby analyses that receive no decision are removed from the calculation. In contrast, this work retains indeterminate outcomes in the denominator. Under the EI convention, the same evaluation would report 99.12 percent sensitivity for ventricular fibrillation rather than 98.21 percent. Consequently, methods reporting similar percentages may not be measuring the same quantity. This distinction is particularly important for cprINSIGHT, which reaches no decision for approximately 30 percent of clinical analyses. Its published performance therefore reflects only the analyses that received a decision rather than the entire clinical cohort.

A second important difference is the degree of independence between the training and test data. Three of the studies in Table~\ref{tab:comparison}, including this work, evaluate performance on a cohort collected by an institution different from the one providing the training data. Among these studies, only the present work additionally enforces patient-level partitioning of the electrocardiograms and repeats the train-validation split generation process. The remaining studies either do not report their separation procedures or, in one case, reuse all compression artifacts for both training and testing. Consequently, superficially similar performance figures may not reflect the same level of protection against data leakage.

A third difference is the operating point at which the comparison is made. Among the studies that report a window length, the analysis window ranges from 8 to 16 seconds. The proposed cascade satisfies both adjudicated AHA criteria using an eight-second window while requiring only 6.68~ms per classification on a single CPU thread. The only other method operating at eight seconds assumes mechanical compressions at a fixed and known rate and reports pooled sensitivity and specificity rather than the rhythm-specific metrics used here. The two methods achieving broadly comparable rhythm-level performance without an auxiliary sensor both operate on fourteen-second windows, and one reports a latency of 1.1~s per decision.


Expressed in clinical terms, the observed errors remain infrequent at the 0~dB operating condition. Across the six evaluated noise scenarios, between two and four subjects per 100 with coarse ventricular fibrillation are classified as non-shockable, while fewer than one sinus-rhythm subject per 100 is classified as shockable. Performance deteriorates at a compression-artifact level of $-3$~dB. Under this condition, approximately one sinus-rhythm subject in 70 is classified as shockable instead of roughly one in 150 at 0~dB, and approximately one subject in 18 from the other non-shockable rhythms is misclassified rather than one in 24.

\subsection{Implications for Public-Access AED Platforms}
\label{sec:implications}

The requirements stated in Section~\ref{sec:intro} were derived from the setting in which an AED operates rather than from the classification task alone, and the results can be read against each of them. The cascade uses only the ECG channel that the device already acquires, so rhythm analysis during compressions adds no reference channel, sensor, or associated hardware cost to units that may remain installed for years between uses and are operated by lay rescuers \cite{hazinski2005lay}. Each decision requires two forward passes and approximately 103 million multiply-accumulate operations, and the eight-bit model occupies 1.45~MiB without changing any compliance verdict (Section~\ref{sec:deployment}). The adjudicable AHA criteria are satisfied at a compression signal-to-artifact ratio of 0~dB. Performance degrades at $-3$~dB, indicating that the available performance margin narrows under heavier compression artifacts. Taken together, these results show that rhythm analysis during compressions can be approached as a change to the signal-processing software of the device rather than to its sensing hardware.

The classifier is only one component of such a change. A defibrillator that analyzes the rhythm during compressions must also define how each classifier output is translated into the prompts given to the rescuer, and this study does not evaluate that behavior. One possible integration path is to retain the conventional pause for analysis as a fallback, so that a segment declared indeterminate leads to a brief interruption of compressions while segments receiving a definitive classification do not. Across the six conditions at 0~dB, this fallback would be invoked for 1.07 to 1.48 percent of segments, but the resulting reduction in hands-off time would depend on the analysis schedule and prompt sequence of the device and would need to be measured in that context. Any such modification would also remain subject to regulatory review and to the performance requirements of IEC 60601-2-4 \cite{iec60601_2_4} before it could be introduced in devices already in service.

The quantization experiments also carry a practical lesson for deploying compact models of this kind. The composition of the calibration set had a larger effect than the choice of quantization scheme. When calibration samples followed the natural corpus distribution, only four of the 4{,}396 samples corresponded to asystole, reflecting the scarcity of asystole in the training corpus noted in Section~\ref{sec:augmentation}, and the quantized cascade lost 6.5 percentage points of asystole specificity while all other metrics remained essentially unchanged. Drawing equal numbers of samples from each rhythm class and adding the synthetic asystole examples used during training restored asystole specificity to the single-precision level. Calibration data for a rhythm classifier should therefore reflect the rhythms and signal characteristics expected in use rather than the class distribution of the available corpus, a consideration that applies whenever a model is compressed for execution on a device with limited resources \cite{jhaveri2024tinyml}.

\subsection{Limitations}
\label{sec:limitations}

The first limitation concerns the construction of the evaluation signals, which remain synthetic despite the use of clinical compression artifacts. Clean ECG recordings are combined with separately recorded compression artifacts rather than being recorded during chest compressions themselves. Although the artifacts were acquired during real resuscitation events and the mixing protocol follows the companion study, the resulting signals remain an approximation of CPR-corrupted ECG. In particular, additive mixing cannot reproduce the full range of physiological and mechanical interactions that may occur when cardiac activity and chest compressions are recorded simultaneously. The reported results should therefore be interpreted as performance under controlled contamination rather than as a direct measure of performance during clinical CPR.

The second limitation concerns the coverage of asystole and rapid supraventricular tachycardia in the test cohort. The cohort contains only nine asystole subjects and 21 rapid supraventricular tachycardia subjects. For asystole, the standard deviation of the cascade across the 15 train-validation splits ranges from 10.2 to 13.0 percentage points across the twelve conditions, indicating substantial uncertainty in the measured rate. The available sample is sufficient to demonstrate that the classifier no longer fails catastrophically on this rhythm, but not to establish a reliable level of performance. The difficulty of asystole classification during chest compressions is not unique to this work. A clinical study of an AED algorithm that analyzes ECG during compressions reported 86.5 percent specificity for asystole and identified low-amplitude ECG as a persistent challenge \cite{didon2024clinical}. That system achieves its highest overall performance by pausing compressions and performing a confirmation analysis on a cleaner signal, whereas the classifier studied here operates directly on compression-contaminated ECG.

The third limitation concerns differences in rhythm composition between the training and test cohorts. Other non-shockable rhythms, premature ventricular contractions, and rapid supraventricular tachycardia together account for 113 of the 392 non-shockable subjects in the test cohort, yet no corresponding training classes carry those labels. Performance on these rhythms therefore depends on the ability of the classifier to generalize beyond the rhythm categories encountered during training. For this reason, rhythm-level specificities are reported separately throughout the paper. Doing so makes it possible to distinguish performance on included rhythm classes from performance on rhythm classes that were effectively unseen during training.

A final limitation concerns deployment validation. The arithmetic, memory, and latency figures reported in this study were obtained on a general-purpose processor, and the cascade has not been measured on defibrillator hardware. Deployment on a specific AED platform would therefore require independent measurement of latency and memory use, together with verification of classification performance after quantization, on that platform.

%% file: 08_conclusion_R2.tex
\section{Conclusion}
\label{sec:conclusion}

This work demonstrates that reliable shockable-rhythm classification during ongoing CPR can be achieved directly from compression-contaminated ECG, without artifact removal and without auxiliary sensing signals. To address the asymmetric performance requirements of AED rhythm analysis, the proposed method employs a cascade of three convolutional networks trained using different class ratios. A balanced network generates the initial classification, which is then verified by a network biased toward the proposed outcome. Disagreement is treated as an indication of insufficient evidence and results in an indeterminate outcome.

At an eight-second analysis window, the cascade satisfies both adjudicable AHA criteria across all six evaluated noise conditions at a compression signal-to-artifact ratio of 0~dB. Performance degrades at $-3$~dB, where the classifier no longer satisfies the two adjudicated specificity criteria. Nevertheless, sensitivity for both shockable rhythms remains above the corresponding AHA requirements. The proposed cascade requires only 6.68~ms per decision on a single CPU thread, and eight-bit quantization reduces the model size to 1.45~MiB without changing any compliance verdict. These figures, obtained on a general-purpose processor, indicate that the computational requirements of the method are modest relative to the analysis interval.

The principal limitation of this work is that the compression contamination is synthetic. Although the artifacts were recorded during real resuscitation events, additive mixing cannot fully reproduce the physiological and mechanical interactions present during simultaneous chest compressions and cardiac activity. The reported results therefore reflect performance under controlled contamination rather than clinical CPR conditions. Future work should evaluate the proposed cascade on recordings obtained during actual chest compressions. Confirming the present findings in that setting is an important step toward clinical validation. Nevertheless, the present results demonstrate that reliable rhythm analysis during ongoing CPR is feasible without artifact removal or auxiliary sensors, which supports pursuing it as a change to the signal-processing software of AEDs rather than to their sensing hardware.

\section*{Data and Code Availability}

The clean ECG recordings used for training are publicly available from PhysioNet
\cite{goldberger2000physiobank}. The CPR artifact recordings and the test cohort
were provided by Defibtech, LLC under a research agreement and are not publicly
distributable. The trained model weights and preprocessing scripts are available
from the corresponding author upon reasonable request. 